\documentclass[useAMS, usenatbib,usegraphicx]{mn2e}
\usepackage{times, subfigure}
\title[ HerMES: SPIRE observations of LBGs]{HerMES: Herschel-SPIRE 
observations of Lyman Break Galaxies}
\author[D.~Rigopoulou et al.]
{\parbox{\textwidth}{D.~Rigopoulou,$^{1,2}$\thanks{E-mail: \texttt{d.rigopoulou1@physics.ox.ac.uk}}
G.~Magdis,$^{3}$
R.J.~Ivison,$^{4,5}$
A.~Amblard,$^{6}$
V.~Arumugam,$^{5}$
H.~Aussel,$^{3}$
A.~Blain,$^{7}$
J.~Bock,$^{7,8}$
A.~Boselli,$^{9}$
V.~Buat,$^{9}$
D.~Burgarella,$^{9}$
N.~Castro-Rodr{\'\i}guez,$^{10,11}$
A.~Cava,$^{10,11}$
P.~Chanial,$^{12}$
D.L.~Clements,$^{12}$
A.~Conley,$^{13}$
L.~Conversi,$^{14}$
A.~Cooray,$^{6,7}$
C.D.~Dowell,$^{7,8}$
E.~Dwek,$^{15}$
S.~Eales,$^{16}$
D.~Elbaz,$^{3}$
D.~Farrah,$^{17}$
A.~Franceschini,$^{18}$
J.~Glenn,$^{13}$
M.~Griffin,$^{16}$
M.~Halpern,$^{19}$
E.~Hatziminaoglou,$^{20}$
J.-S.~Huang,$^{21}$
E.~Ibar,$^{4}$
K.~Isaak,$^{16}$
G.~Lagache,$^{22}$
L.~Levenson,$^{7,8}$
N.~Lu,$^{7,23}$
S.~Madden,$^{3}$
B.~Maffei,$^{24}$
G.~Mainetti,$^{18}$
L.~Marchetti,$^{18}$
H.T.~Nguyen,$^{8,7}$
B.~O'Halloran,$^{12}$
S.J.~Oliver,$^{17}$
A.~Omont,$^{25}$
M.J.~Page,$^{26}$
P.~Panuzzo,$^{3}$
A.~Papageorgiou,$^{16}$
C.P.~Pearson,$^{1,27}$
I.~P{\'e}rez-Fournon,$^{10,11}$
M.~Pohlen,$^{16}$
D.~Rizzo,$^{12}$
I.G.~Roseboom,$^{17}$
M.~Rowan-Robinson,$^{12}$
B.~Schulz,$^{7,23}$
Douglas~Scott,$^{19}$
N.~Seymour,$^{26}$
D.L.~Shupe,$^{7,23}$
A.J.~Smith,$^{17}$
J.A.~Stevens,$^{28}$
M.~Symeonidis,$^{26}$
M.~Trichas,$^{12}$
K.E.~Tugwell,$^{26}$
M.~Vaccari,$^{18}$
I.~Valtchanov,$^{14}$
L.~Vigroux,$^{25}$
L.~Wang,$^{17}$
G.~Wright,$^{4}$
C.K.~Xu$^{7,23}$ and
M.~Zemcov$^{7,8}$}\vspace{0.4cm}\\
\parbox{\textwidth}{$^{1}$Space Science \& Technology Department, Rutherford Appleton Laboratory, Chilton, Didcot, Oxfordshire OX11 0QX, UK\\
$^{2}$Astrophysics, University of Oxford, Keble Road, Oxford OX1 3RH, UK\\
$^{3}$Laboratoire AIM-Paris-Saclay, CEA/DSM/Irfu - CNRS - Universit\'e Paris Diderot, CE-Saclay, pt courrier 131, F-91191 Gif-sur-Yvette, France\\
$^{4}$UK Astronomy Technology Centre, Royal Observatory, Blackford Hill, Edinburgh EH9 3HJ, UK\\
$^{5}$Institute for Astronomy, University of Edinburgh, Royal Observatory, Blackford Hill, Edinburgh EH9 3HJ, UK\\
$^{6}$Dept. of Physics \& Astronomy, University of California, Irvine, CA 92697, USA\\
$^{7}$California Institute of Technology, 1200 E. California Blvd., Pasadena, CA 91125, USA\\
$^{8}$Jet Propulsion Laboratory, 4800 Oak Grove Drive, Pasadena, CA 91109, USA\\
$^{9}$Laboratoire d'Astrophysique de Marseille, OAMP, Universit\'e Aix-marseille, CNRS, 38 rue Fr\'ed\'eric Joliot-Curie, 13388 Marseille cedex 13, France\\
$^{10}$Instituto de Astrof{\'\i}sica de Canarias (IAC), E-38200 La Laguna, Tenerife, Spain\\
$^{11}$Departamento de Astrof{\'\i}sica, Universidad de La Laguna (ULL), E-38205 La Laguna, Tenerife, Spain\\
$^{12}$Astrophysics Group, Imperial College London, Blackett Laboratory, Prince Consort Road, London SW7 2AZ, UK\\
$^{13}$Dept. of Astrophysical and Planetary Sciences, CASA 389-UCB, University of Colorado, Boulder, CO 80309, USA\\
$^{14}$Herschel Science Centre, European Space Astronomy Centre, Villanueva de la Ca\~nada, 28691 Madrid, Spain\\
$^{15}$Observational  Cosmology Lab, Code 665, NASA Goddard Space Flight  Center, Greenbelt, MD 20771, USA\\
$^{16}$Cardiff School of Physics and Astronomy, Cardiff University, Queens Buildings, The Parade, Cardiff CF24 3AA, UK\\
$^{17}$Astronomy Centre, Dept. of Physics \& Astronomy, University of Sussex, Brighton BN1 9QH, UK\\
$^{18}$Dipartimento di Astronomia, Universit\`{a} di Padova, vicolo Osservatorio, 3, 35122 Padova, Italy\\
$^{19}$Department of Physics \& Astronomy, University of British Columbia, 6224 Agricultural Road, Vancouver, BC V6T~1Z1, Canada\\
$^{20}$ESO, Karl-Schwarzschild-Str. 2, 85748 Garching bei M\"unchen, Germany\\
$^{21}$Harvard-Smithsonian Center for Astrophysics, MS65, 60 Garden Street,  Cambridge,  MA02138, USA\\
$^{22}$Institut d'Astrophysique Spatiale (IAS), b\^atiment 121, Universit\'e Paris-Sud 11 and CNRS (UMR 8617), 91405 Orsay, France\\
$^{23}$Infrared Processing and Analysis Center, MS 100-22, California Institute of Technology, JPL, Pasadena, CA 91125, USA\\
$^{24}$School of Physics and Astronomy, The University of Manchester, Alan Turing Building, Oxford Road, Manchester M13 9PL, UK\\
$^{25}$Institut d'Astrophysique de Paris, UMR 7095, CNRS, UPMC Univ. Paris 06, 98bis boulevard Arago, F-75014 Paris, France\\
$^{26}$Mullard Space Science Laboratory, University College London, Holmbury St. Mary, Dorking, Surrey RH5 6NT, UK\\
$^{27}$Institute for Space Imaging Science, University of Lethbridge, Lethbridge, Alberta, T1K 3M4, Canada\\
$^{28}$Centre for Astrophysics Research, University of Hertfordshire, College Lane, Hatfield, Hertfordshire AL10 9AB, UK}\vspace{2cm}}

\begin{document}

\date{Accepted 2010 September. Received 2010 September; in original form 2010
June}

\pagerange{\pageref{firstpage}--\pageref{lastpage}} \pubyear{2010}

\maketitle

\label{firstpage}
\begin{abstract}

We present first results of a study of the submillimetre (rest frame
far-infrared) properties of $z\sim$3 Lyman Break Galaxies (LBGs) and
their lower-redshift counterparts BX$/$BM galaxies, based on
{\it Herschel}-SPIRE observations of the Northern field of the Great
Observatories Origins Deep Survey (GOODS-N). We use stacking analysis to
determine the properties of LBGs well below the current limit of the
survey. Although LBGs are not detected individually, stacking the
infrared luminous LBGs (those detected with Spitzer at 24 $\mu$m)
yields a statistically significant submm detection with mean flux
$\langle S_{250} \rangle$= 5.9$\pm$1.4 mJy confirming the power of SPIRE in
detecting UV-selected high-redshift galaxies at submillimetre
wavelengths.  In comparison, the Spitzer 24 $\mu$m detected BX$/$BM
galaxies appear fainter with a stacked value of 
$\langle S_{250} \rangle$ =
2.7$\pm$0.8 mJy. By fitting the Spectral Energy Distributions (SEDs)
we derive median infrared luminosities, $L_{\rm IR}$, of
2.8$\times$10$^{12}$L$_{\odot}$ and 1.5$\times$10$^{11}$L$_{\odot}$
for $z\sim$3 LBGs and BX$/$BMs, respectively. We find that
$L_{\rm IR}$ estimates derived from present measurements are in good
agreement with those based on UV data for z$\sim$2 BX$/$BM galaxies, unlike
the case for $z\sim$3 infrared luminous LBGs where the UV
underestimates the true $L_{\rm \bf {IR}}$. Although sample selection effects 
may influence this result we suggest that differences in physical
properties (such as morphologies, dust distribution and extent of
star-forming regions) between $z\sim$3 LBGs and z$\sim$2 BX$/$BMs 
may also play a significant role.

\end{abstract}


\section{Introduction}

The broadband dropout technique has been a very succesful tool for
discovering high-redshift galaxies, the so-called Lyman break galaxies
(LBGs, e.g. Steidel \& Hamilton 1993, Steidel et al. 1999).  The
initial selection focused on $z\sim$3 samples. The same colour
criteria were later extended to select Lyman break galaxies at
1.4$<$z$<$2.5 (the so called BX$/$BM objects) with approximately the
same range of UV luminosity and intrinsic UV colours as the $z\sim$3
LBGs (Reddy et al. 2006). The dropout broadband technique provides a
complete census of UV light at high redshift, with well over a thousand
galaxies detected at z$>$1.5. Recent detailed studies
including {\it Spitzer} observations have shown that some of these galaxies
have large stellar masses $>$10$^{10}$M$_{\odot}$ (e.g. Rigopoulou et
al. 2006, Reddy et al. 2006, Magdis et al. 2008, 2010a) while their
comoving volume density at $z\sim$3 is $\sim$0.005 Mpc$^{-3}$
(e.g. Reddy \& Steidel 2009).

A number of issues related to the nature and properties of $z\sim$3
LBGs remain unclear. The dust-corrected star formation rate (SFR) of
LBGs can be as high as 100 M$_{\odot}$/yr, which would
correspond to $S_{850} \sim$ 1mJy depending on specific dust
parameters (Chapman et al. 2009). However, the search for the
sub-millimetre (submm) counterparts of LBGs has proven challenging due
to uncertainties in the relations used to predict the rest-frame
far-infrared luminosity from the UV. Peacock et al. (2000) analysed
the submm emission from star forming galaxies with the highest UV
star-formation rates and found that they were statistically detected
with a flux density S$_{850}$=0.2 mJy for a star formation rate of 1
h$^{-2}$ M$_{\odot} / yr$.  Chapman et al. (2000, 2009) reported the
submm detection of Westphal MMD-11 and Westphal-MM8, while Rigopoulou
et al. (2010) reported mm detections of a further two LBGs, EGS-D49
and EGS-M28 selected based on their strong MIPS 24 $\mu$m emission
(e.g. Huang et al. 2005). Despite these promising detections the
properties of the FIR and submm emission from LBGs, their dust content
and their possible contribution to the cosmic far-infrared background
is still largely unconstrained.

With the advent of {\it Herschel} (Pilbratt et
al. 2010)  it is now possible to investigate the submm 
(rest-frame far-infrared) properties of
LBGs. In this letter we report first results on the far-infrared
properties of LBGs
based on observations that are part of the {\it Herschel}
Multi-tiered Extragalactic Survey (HerMES), a Guaranteed Time project 
that will eventually result in a
variety of surveys of varying depth and area which will be
covered in five photometric bands (110, 160, 250, 350, 500 $\mu$m;
Oliver et al. 2010). The results presented here are based on HerMES
data taken as part of the {\it Herschel} Science Demonstration
Phase. Throughout this paper we assume $\Omega_{m}$ = 0.3, 
$\Omega_{\Lambda}$=0.72 and H$_{0}$ = 72 km s$^{-1}$ Mpc$^{-1}$.

\section{Observations, Sample Selection and Analysis}

\subsection{Herschel Observations}

Submm observations of the Northern field of the Great Observatories
Origins Deep Survey (GOODS-N) were carried out at 250, 350 and 500 $\mu$m,
with the Spectral and Photometric Imaging Receiver (SPIRE). The
instrument and its capabilities are described in Griffin et al. (2010),
while the SPIRE astronomical calibration methods and accuracy are
outlined in Swinyard et al. (2010). The GOODS-N images are amongst the
deepest possible with SPIRE and, the instrumental noise is less than
the confusion noise from overlapping faint sources.  Confusion noise
values of 5.8, 6.3 and 6.8 mJy beam$^{-1}$ at 250, 350 and 500 $\mu$m
respectively, are reported in Ngyuen et al. (2010).
Besides blind source extraction resulting in single-band catalogues
(SCAT, see Smith et al. 2010, in prep.), a novel source extraction
method based on 24 $\mu$m priors has been developed to
detect sources as close as possible to the confusion limit (see
Roseboom et al. 2010, hereafter XID catalogue). 
The method uses 
a matrix inversion
technique which relies on the assumption that sources detected in the
250 $\mu$m band will also be detected at 24 $\mu$m deep
surveys (e.g. Marsden et al. 2009). The 24 $\mu$m catalogue
positions are then used to find sources in the Herschel 250 $\mu$m
images. The flux densities of the sources are allowed to vary until finally a
set of flux densities is found that produce the best match to the image.
In the current study we have made use of both blind (SCAT) and 24 $\mu$m
prior source catalogues (XID), while for the stacking analysis we have used
calibrated GOODS-N SPIRE images.

\begin{figure*}[h]
\centering
\subfigure[] 
{
    \label{fig:sub:a}
    \includegraphics[width=60mm,angle=270]{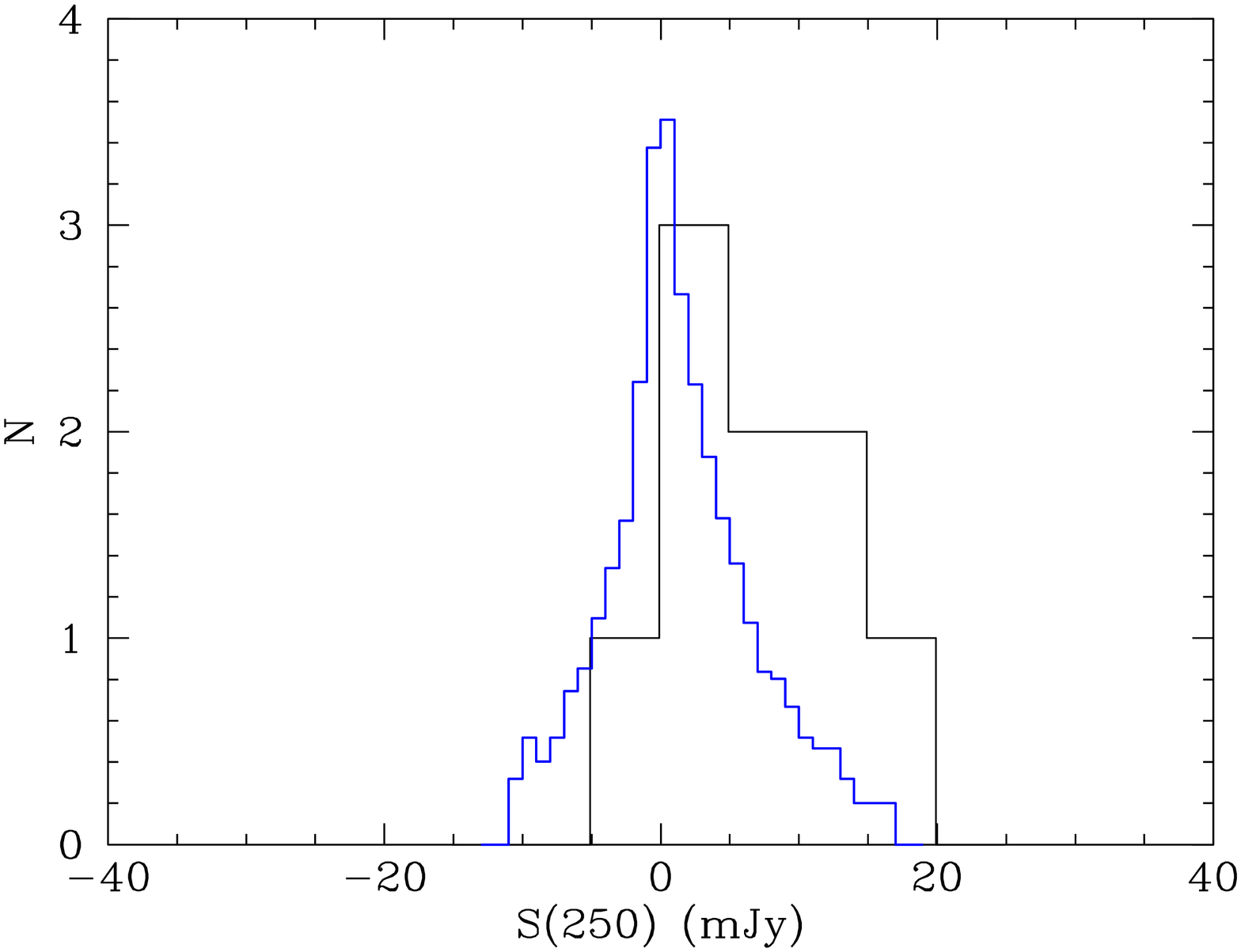}
}
\hspace{0.5cm}
\subfigure[] 
{
    \label{fig:sub:b}
    \includegraphics[width=60mm,angle=270]{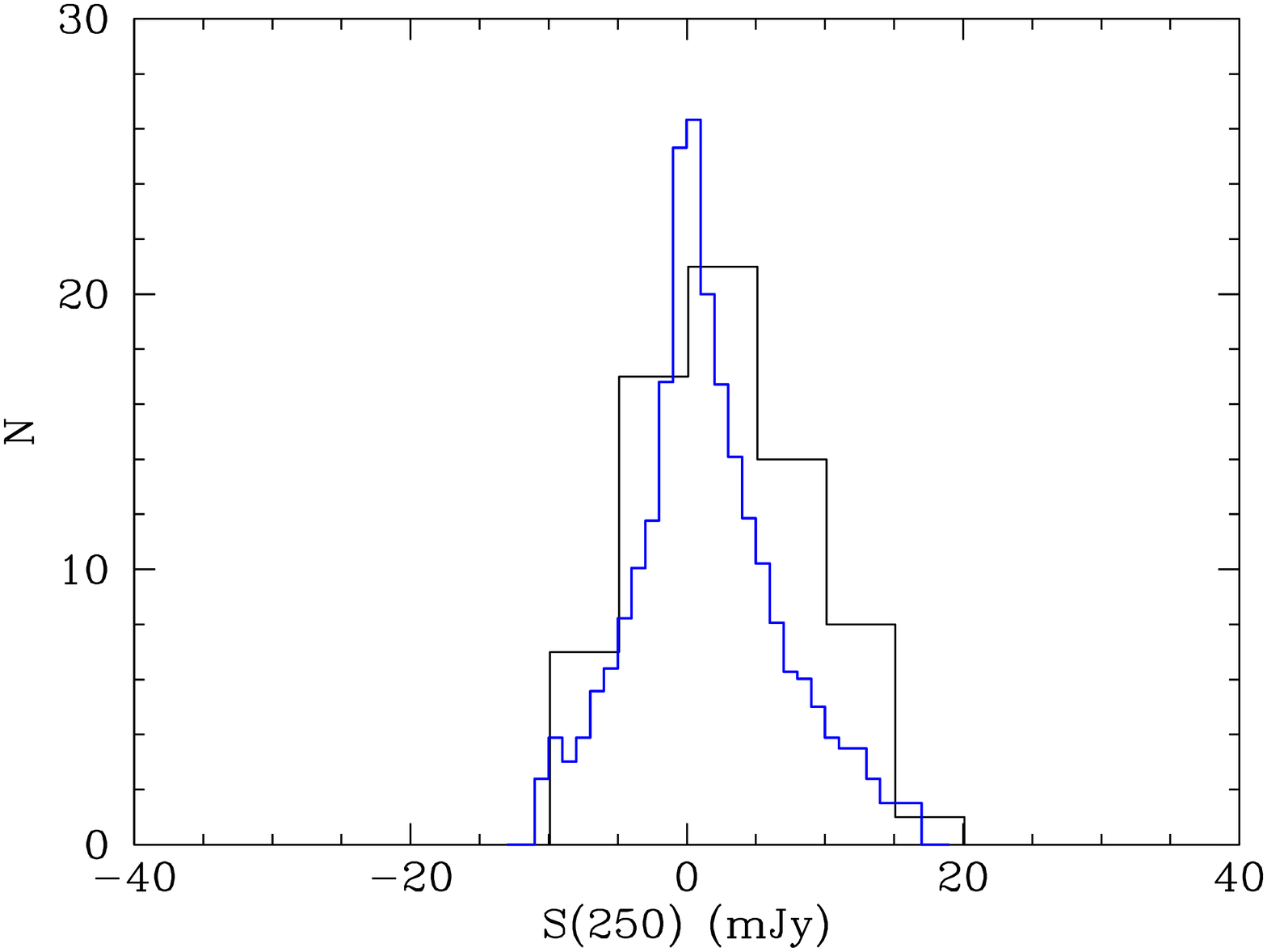}
}
\caption{Histograms showing the 250 $\mu$m flux density distributions for 
LBGs (left) and BX$/$BMs (right). The blue curves show the flux density 
distribution in the entire GOODS-N field (control sample). 
In each case the control sample has been normalised by a factor equal
to the total
number of objects in each case ({\it N}=9 for LBGs, {\it N}=69 for BX$/$BM) divided by the
total number of pixels in the whole GOODS-N image.}

\end{figure*}

\subsection{Sample Selection and Analysis}

The GOODS-N region contains 58 UV selected $z\sim$3 LBGs
(Steidel et al.  2003) and 212 UV selected BX$/$BM objects
(Reddy et al. 2006) down to $R\leq$25.5. 49 LBGs and 200 BX$/$BMs
have been detected with the {\it Spitzer} Infrared Array Camera (IRAC, (down
to 25.0 mag(AB) at 3.6 $\mu$m) and 9 LBGs and 69 BX$/$BMs have also been
detected with the Multi Imager Photometer for {\it Spitzer} (MIPS, down to
$S_{24}$=20 $\mu$Jy, 5$\sigma$). In the current study we investigate
the submm properties of the UV selected LBGs and BX$/$BMs
focusing particularly on the sub-sample with MIPS detections: 9 $z\sim$3
LBGs and 69 1.5$\leq z \leq$2.5 BX$/$BM (hereafter the MIPS-LBG and
MIPS-BX$/$BM samples). All LBG and BX$/$BM galaxies have spectroscopic
redshifts determined from optical spectroscopy (Steidel et al. 2003,
Reddy et al. 2006) which has also been used to confirm the absence of
strong high ionization emission lines indicative of the presence of
AGN. Objects classified as AGN$/$QSO are excluded from this study.

The LBG and BX$/$BM samples were matched to the blind (SCAT) and 
priors (XID) catalogues. An object is considered
detected when its flux is at least 3$\sigma$ above the noise
(confusion plus instrumental).  
None of the LBGs and only three of the BX$/$BM objects are detected
in the HerMES GOODS-N priors catalogue down to $S_{250} \sim$20 mJy. 
We discuss the properties of individual sources in section 2.3.

To assess the reliability of SPIRE detections at faint flux levels
($\leq$20 mJy) close to the confusion limit we first look at the flux
distribution (per pixel) of the whole GOODS-N map. For the present discussion
we focus on the 250 $\mu$m band which is the most sensitive and
has the smallest beamsize (18$^{\prime \prime}$).  We find that the fraction of pixels
above 5, 10 and 20 mJy is 6, 4 and 0.6 percent, respectively. These
percentages imply that 16(4.5), 10(3) and 1.5(0.23) sources out of the 270(78)
sources studied here could
be associated with spurious detections at 5, 10 and 20 mJy. The number
in parentheses correspond to the MIPS detected objects. 

Since none of the LBGs and only three of the BX$/$BMs
are detected individually, we examine the properties of the two samples
via stacking analysis. 
We measure flux densities directly from the
calibrated GOODS-N SPIRE 250, 350 and 500 $\mu$m images 
(at the optical positions of the LBGs and BX$/$BMs). 
For the stacking we
first consider the optically-selected LBG and BX$/$BM samples. 
For the analysis we
employ median stacking and stack at the optical position for each object.
We exclude sources near bright objects (within 18$^{\prime \prime}$ at 250 $\mu$m) to avoid
contamination of the measured signal. Stacking the 
UV selected samples returned no detections in any of the 
three SPIRE bands. For the
LBGs we determine 3$\sigma$ upper limits of $S_{250}<$2.8 mJy, 
$S_{350}<$1.5mJy and $S_{500}<$0.9mJy. For the BX$/$BMs the corresponding
(3$\sigma$) upper limits are $S_{250}<$2.02 mJy, 
$S_{350}<$1.2mJy and $S_{500}<$0.6mJy, respectively.
\begin{figure*}[h]
\centering
\subfigure[] 
{
    \label{fig:sub:a}
    \includegraphics[width=60mm,angle=270]{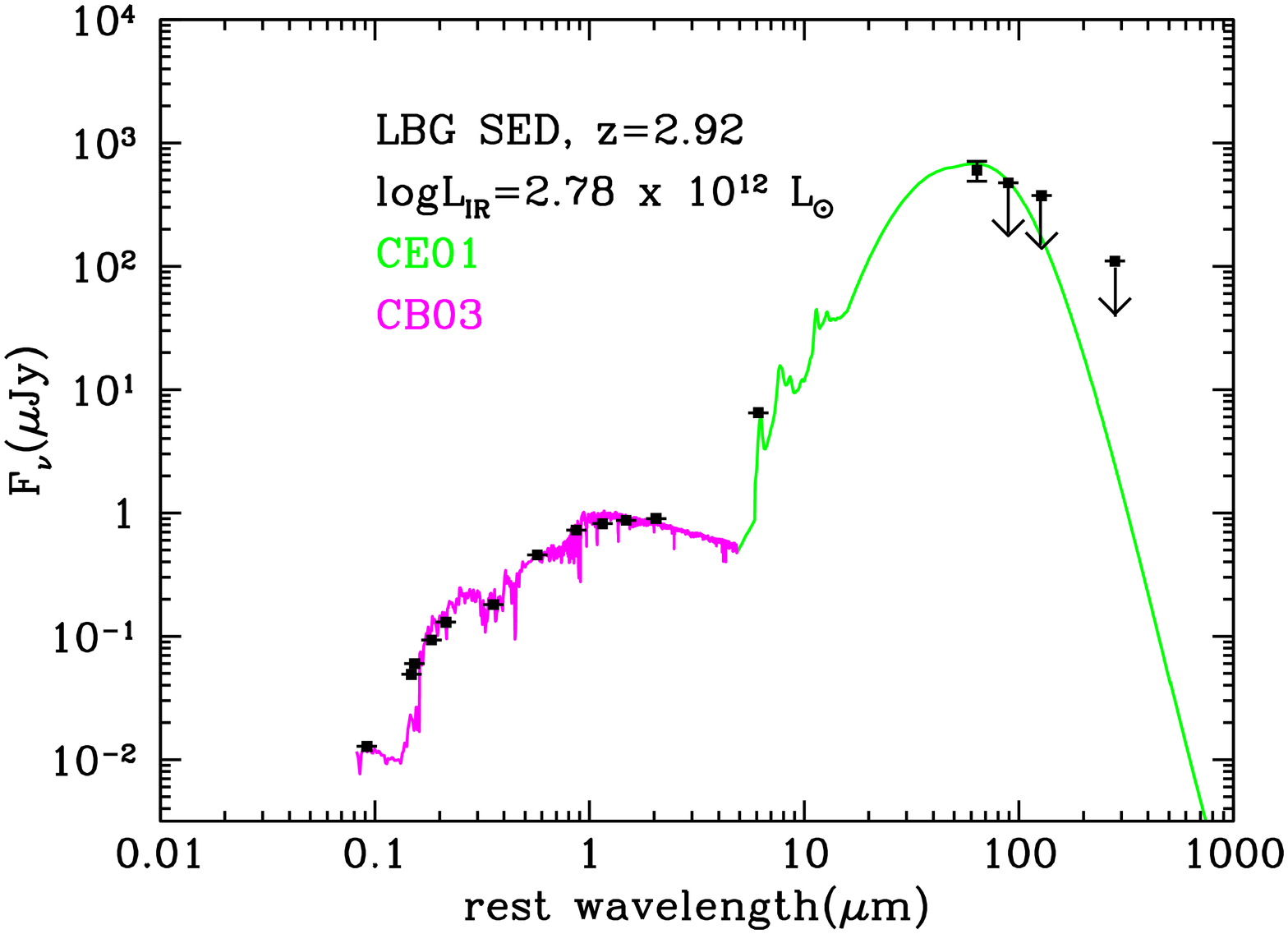}
}
\hspace{0.5cm}
\subfigure[] 
{
    \label{fig:sub:b}
    \includegraphics[width=60mm,angle=270]{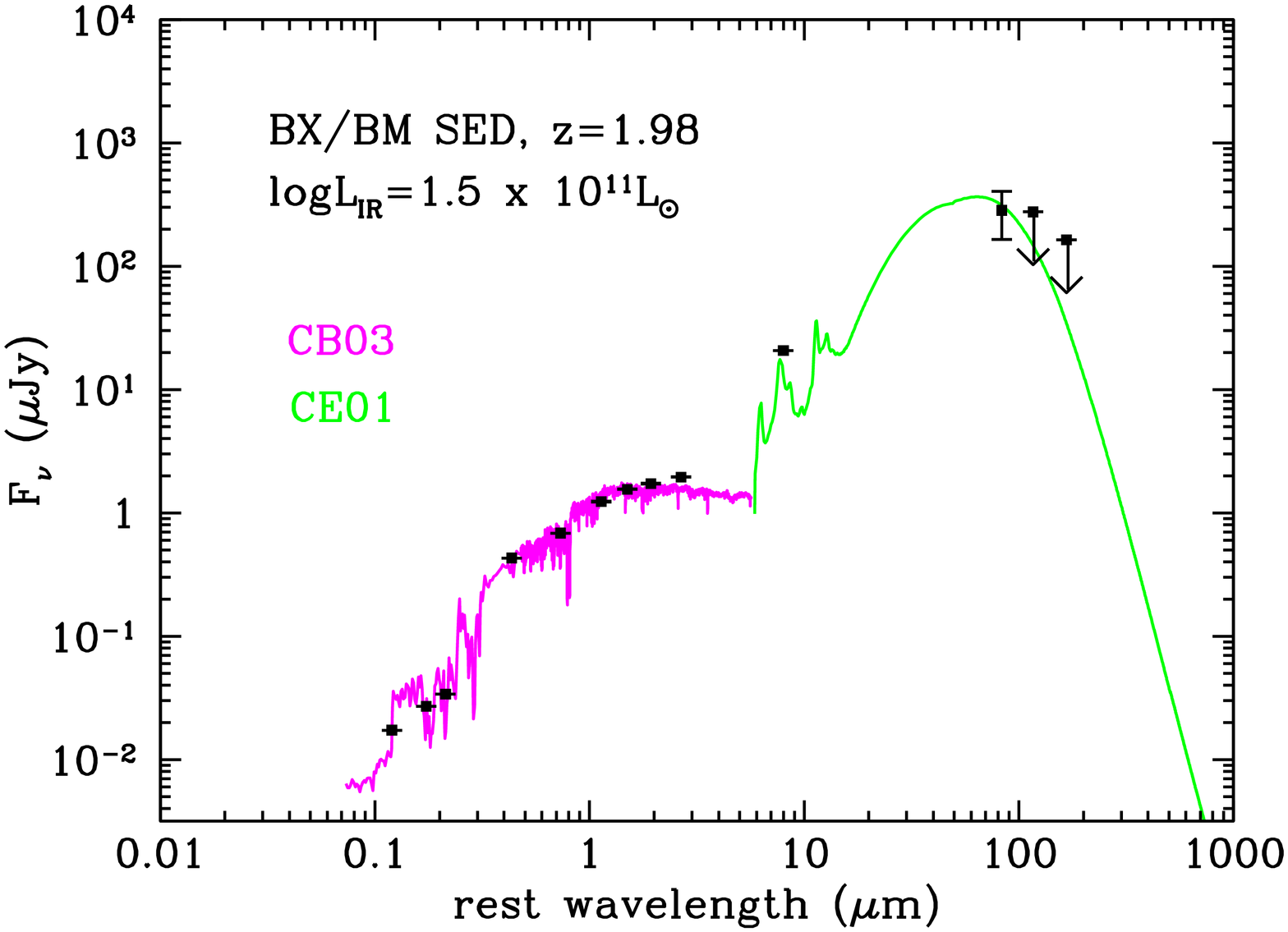}
}

\caption{Rest frame average SED of MIPS-detected LBGs (left) and 1$<z<$2.5 
BX$/$BMs (right) galaxies. For the SEDs we used mean values of {\it UGRViJK}, 
IRAC, MIPS, the mean value derived from SPIRE 250$\mu$m and upper limits 
from SPIRE
350, 500 $\mu$m and 
Aztec 1.1 mm measurements (for $z\sim$3 LBGs only). The rest-frame UV$/$optical
is fit with BC03 models (magenta) while the mid$/$far-infrared part of the 
SED is fit with CE01 templates (green line).}
\end{figure*}

The same stacking technique was employed to investigate the properties
of the MIPS detected LBGs and BX$/$BMs. In Figure 1 we show the
histograms of the 250 $\mu$m flux density distributions for the
MIPS-LBGS and MIPS-BX$/$BMs. In the same plot, we show the normalised
distribution of flux densities per pixel for the whole GOODS-N
image. In both cases the distributions have positive skews and in the
case of LBGs there is a clear positive tail implying that pixels
containing flux from the LBGs have higher flux than the average pixel
in the map. The case for the BX$/$BMs is not as clear. Although the
BX$/$BM sample is bigger (69 objects vs 9) the population as a whole
does not appear to strongly emit in the submm.  In order to
confirm that the two distributions are in fact different, we carry out
a K-S two-sample test. The test results, {\it D}= 0.31 and (probability)
$\alpha$({\it D}) = 0.03, suggest that the two flux density distributions
are intrinsically different at the 2$\sigma$ level.  The mean flux
densities are $\langle S_{250} \rangle$ = 5.9$\pm$1.4 mJy (LBG) and
$\langle S_{250} \rangle$ =2.7$\pm$0.8 mJy (BX$/$BM). The errors reported 
in these
measurements have been quantified by stacking at 9 (for the LBGs) and
69 (for the BX$/$BMs) random positions and then repeating the process
40,000 times. The 1$\sigma$ value of the derived distribution is adopted as
the uncertainty of the measurement. Stacking at 350 and 500 $\mu$m
resulted in no formal detections. Instead we quote 3$\sigma$ upper
limits of $S_{350}<$ 4.9 mJy and $S_{500}<$ 3.4 mJy for the MIPS-LBGs and
$S_{350}<$ 2.6 mJy and $S_{500}<$ 1.6mJy for the MIPS-BX$/$BMs, respectively.

A likely physical explanation for this difference in detection rates comes
from the MIR properties of the two samples. For $z\sim$3 LBGs the MIPS
24$\mu$m band contains contributions from 6$\mu$m (hot) dust continuum
plus the 6.2 $\mu$m emission from Polycyclic Aromatic Hydrocarbons (PAHs).
The PAH contribution to the MIPS 24$\mu$m flux is $\sim$20--30 percent
(e.g. Huang et al. 2005), therefore, the MIPS 24 $\mu$m flux mostly reflects
the strength of the underlying continuum. Thus, the MIPS detected
LBGs (with the strongest 6 $\mu$m continuum) are likely to be amongst the
most luminous $z\sim$3 LBGs.
For z$\sim$2 BX$/$BMs the MIPS
24$\mu$m band includes contributions from both the 7.7$\mu$m PAH
feature and the underlying continuum. However, z$\sim$2 galaxy
populations show a wide variety in their $L_{5-8.5 \mu m}$ rest-frame
luminosity, with BX$/$BMs, in particular showing a relatively weak MIR
continuum (Reddy 2006b). PAHs in BX$/$BMs, might be instrinsically
weak since metallicity is known to affect their strength (Houck et
al. 2005). The combination of low MIR continuum plus weaker PAH
features may be responsible for the low detection rate in the submm of
24 $\mu$m selected BX$/$BMs. We thus conclude that, $z\sim$3 LBGs detected
by MIPS are likely to be on average more luminous than MIPS detected 
$z\sim$2 BX$/$BMs.

\subsection{Individual detections}

In the previous section we examined the average submm properties of
LBGs and BX$/$BMs here we take a closer look at individual detections,
focusing on objects with MIPS detections. As discussed in section 2.1
of the 69 MIPS-BX$/$BM objects, 3 are detected with $S_{250} >$ 20mJy
and S$/$N $>$ 3 however, two of those, BX1296 and BX1223 lie close
(within $\leq$ 3$^{\prime\prime}$) to bright submm galaxies, BX1296 to
GN07 and BX1223 to GN06 (the notation is from Pope et al. 2005). These
confused cases have not been considered in this work. BM1326 is
clearly detected, with $S_{250}$=22$\pm$5 mJy while a further 3
BX$/$BMs appear in the 250 SPIRE map (and prior-based catalogue) but
with 250 $\mu$m flux densities $<$10 mJy. Additionally, 2 (of the 9)
LBGs, HDFN-M18 and HDFN-M23 appear to be present in the maps (and
prior-based catalogue) although with fluxes below the 10mJy level. We
note that, HDFN-M23 is included in the 5$\sigma$ radio catalogue of
Morrison et al. (2010), with a flux density 21.2$\pm$4 $\mu$Jy.

\section{Results}

\subsection{Spectral Energy Distributions of LBGs}

Fig. 2 shows the average Spectral Energy Distribution (SED) of
MIPS-detected LBGs and BX$/$BM galaxies. The SEDs have been
constructed using available `averaged' {\it UGRViJK}, IRAC, MIPS and SPIRE
flux measurements. For the MIPS-LBGs we also use the 1.1 mm Aztec
measurement from Magdis et al (2010b). We fit the optical$/$near-infrared
part with model SEDs generated using the Bruzual \& Charlot (2003, BC03) 
code, while the mid-to-far infrared part is fit using Chary \&
Elbaz (2001, CE01) template SEDs. In brief, we use BC03 and construct
stellar population models with a Salpeter IMF and constant star
formation rate, which has been shown (van Dokkum et al. 2004, Rigopoulou et
al. 2006, Lai et al. 2007) to provide an adequate description of the
properties of high redshift galaxies with ongoing star formation. Age,
stellar mass, dust reddening {\it E(B-V)} and star formation rates are then
derived from the model fits. It is beyond the scope of the present
work to discuss these results, a detailed analysis of the properties
of the stellar population in {\it Spitzer} detected LBGs can be found in
e.g. Rigopoulou et al. (2006), Magdis et al. (2010a, for LBGs) and Reddy
et al. (2006, for BX$/$BMs). It is however, worth noting the 
differences in the optical part of the SED with the BX$/$BM galaxies
showing a much `bluer' SED.

We fit the far-IR$/$submm part with templates from the CE01 library, with the 
best-fit templates rendering mean 
$\langle L_{\rm IR} \rangle$ values of 2.8($\pm$0.6)$ \times 10^{12}$ L$_{\odot}$ for MIPS-LBGs and
1.5($\pm$0.5)$ \times 10^{11}$ L$_{\odot}$ for MIPS-BX$/$BMs. 
The derived averaged $L_{\rm IR}$ for LBGs is typical of those seen in 
Ultraluminous Infrared Galaxies (ULIRGs). Using the infrared luminosities we
derive average Star Formation Rates $\langle SFRs \rangle$ of 
296M$_{\odot}/$yr and  245M$_{\odot}/$yr, for the $z\sim$3 LBGs and BX$/$BM 
galaxies, respectively. The $<$SFR$>$ derived from the IR for LBGs is in 
agreement with the radio SFR estimate (280$\pm$85M$_{\odot}$ yr$^{-1}$) 
but higher
than the UV SFR estimate (250$^{+35}_{-80}$ M$_{\odot}$ yr$^{-1}$)
reported in Magdis et al (2010b). Turning to the BX$/$BM galaxies, the present
SFR estimates agree well with those derived from the UV (Reddy et al. 2006)
for UGR-selected galaxies.

\subsection{Dust, temperature and mass}

To derive the dust temperature, we use a single temperature greybody fitting
function (Hildebrand 1983) in
which the thermal dust spectrum is approximated by:
$F_{\nu} =Q_{\nu}B_{\nu}(T_{\rm d})$, where $B_{\nu}$ is the Planck function,
$Q_{\nu}=Q_{0}({\nu} / {\nu}_{0})^{\beta_{d}}$ is the dust emissivity (with
1$\leq {\beta_{d}} \leq$2) and,
$T_{\rm d}$ is the effective dust temperature.
For $h\nu / kT_{\rm d} \geq$1 the spectrum becomes :
\begin{equation}
F_{\nu} \propto \frac{\nu^{3+\beta_{d}}}{exp(h\nu/kT_{d})-1}
\end{equation}
Since T$_{d}$ and $\beta_{d}$ are degenerate for sparsely sampled SEDs 
we have fixed $\beta_{d}$ = 1.5 (e.g. Blain et al. 2003)
which is consistent with SED fitting of low and 
high-z systems (e.g. Dunne et al. 2001).  A higher value of $\beta_{d}$ will 
result in lower dust temperatures (Sajina et al. 2006). 
The dust temperature for BM1326, was obtained from the best fit model derived
from minimization of the $\chi^{2}$ values. The uncertainty in the
measurement was obtained by repeating the procedure based on perturbed
values of the photometric points within their errors. 
\begin{figure}
 \includegraphics[width=60mm, angle=-90]{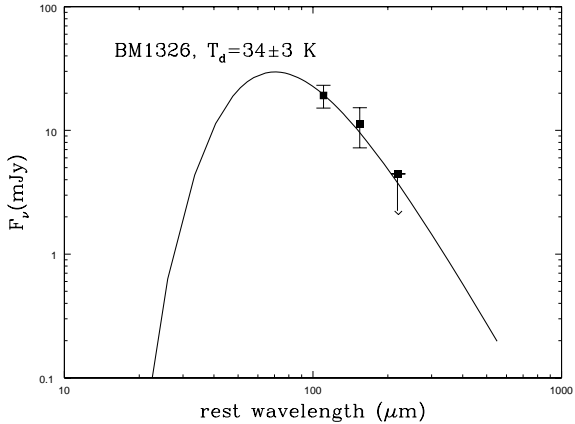} 
 \caption{ Modified blackbody temperature fit for the galaxy 
BM1326 (which is individually detected at the SPIRE bands). The
black filled squares represent the individual detections at 250, 350 
and 500 $\mu$m. For the fit we have fixed $\beta_{\rm d}$=1.5.}
\end{figure}
To derive dust masses we follow:
\begin{equation}
M_{d}=\frac{S_{\nu}
D^{2}_{L}}{\kappa ({\lambda}_{rest})B_{\nu}({\lambda}_{rest},T_{d})}
\end{equation}
where
$M_{d}$ is the total dust mass, $S_{\nu}$ the observed flux density,
$D_{L}$ luminosity distance, $\kappa (\lambda_{rest}$) is the rest frame
dust mass absorption coefficient (taken from Weingartner \& Draine
2001) and B$_{\nu}({\lambda}_{rest}, T_{d})$ is the Planck function.
For the $z\sim$3 LBGs we assume $T_{d}$ = 45 K, a value chosen from  
$T_{d}$ estimates for local ULIRGs (Lisenfeld, Isaak, Hills et al. 1997)
since the average MIPS-LBG appears to 
have L$_{IR}>$ 10$^{12}$L$_{\odot}$. 
We derive dust masses of $M_{d}$=5.5$\pm$1.6$\times$10$^{8}$
M$_{\odot}$ and $M_{d}$=12.8$\pm$2.3$\times$10$^{8}$ M$_{\odot}$ for the
LBGs and for BM1326, respectively.


\section{Dust obscuration in UV selected galaxies}

The present SPIRE observations allow us to
probe the cold dust peak of LBGs, determine their far-IR luminosity,
dust temperature and dust mass from the far-IR alone with minimal additional
assumptions.  
Earlier attempts to detect submm emission from LBGs (with targets 
selected mostly based on SFR(UV) estimates) were not met with 
success (e.g. Chapman et al. 2000, Peacock et al. 2000). These initial 
results led to suggestions
that either $T_{d}$ is high ($T_{d}\geq $90 K) or, that estimates of
$L_{\rm IR}$ from the rest-frame UV and$/$or from the
scatter in the UV-slope/far-IR relation are uncertain (e.g. Chapman
et al 2000). Recently, Rigopoulou et al. (2010) reported 1.2 mm
detections of 2 LBGs in the Extended Groth Strip (EGS), both 
are detected in the MIPS 24 $\mu$m imaging survey of the EGS (for the full
SEDs see Rigopoulou et al 2006). Briefly, their properties are similar
to those of the GOODS-N $z\sim$3 LBGs with $S_{24}$ in 
the 50--100 $\mu$Jy range. Using CE01 models we infer infrared 
luminosities,
L$_{IR} \sim$ few$\times$10$^{12}$ L$_{\odot}$ for each of these
LBGs.

Let us now focus on the MIPS-detected LBGs and BX$/$BMs 
using the mean LBG properties reported in section 3.1 and compare
$L_{\rm IR}$ values from the SPIRE data to those derived from the UV.
L$_{IR, UV}$ is determined as follows: at $z\sim$3, {\it G} and {\it R} bands
correspond to rest-frame 1200 {\AA}   and 1500 {\AA}, respectively, 
thus alowing
us to estimate the slope $\beta$. Assuming solar metallicity, Salpeter
IMF and continuous dust-free star formation models we use the BC03 code
to generate SEDs to fit each of the LBGs,
assuming the Calzetti (2000) attenuation law (but see also Buat et al.
2010 for a discussion of alternative extinction laws). Based on the
best-fit model we derive extinction values {\it E(B-V)} and infer the
observed and intrinsic 1500 {\AA}  flux density and subsequently,
$L_{1500}$ luminosity. We repeat the same process for the two
galaxies with mm detections and the z$\sim$2 mean BX$/$BM galaxies
(using the {\it B} band flux density to estimate the intrinsic 
$L_{1500}$ luminosity). 

The results are plotted in Fig. 4. 
\begin{figure}
\includegraphics[width=60mm, angle=-90]{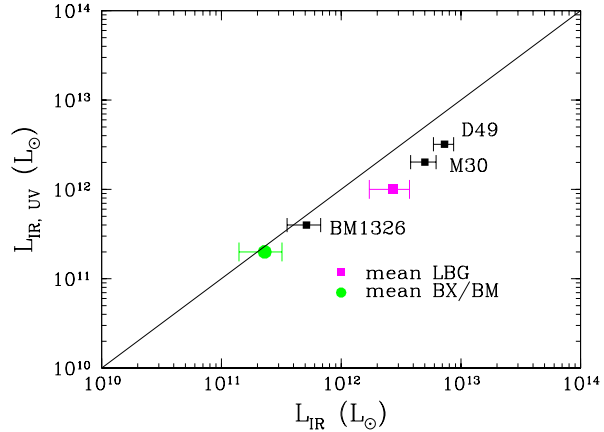} 
\caption{A comparison of estimates of $L_{\rm IR}$ from the present
submm observations and from the UV ($L_{IR, UV}$ for the average
$z\sim$3 LBGs (magenta), the average $z\sim$2 BX$/$BMs (green), together with 
the BM1326 and D49 and M30 (the two LBGs with 1.2 mm detections).
The solid line represents the $L_{IR, UV} \sim$ $L_{\rm IR}$. }
\end{figure}
The UV appears to underestimate the $L_{IR, UV}$ of both the averaged
$z\sim$3 LBG (and the two LBGs with additional mm detections) 
by a factor $\sim$2. This is perhaps not surprising given 
that the LBGs that appear to be detected in the submm regime are all 
ULIRGs. It is known that the 
UV underpredicts the $L_{\rm IR}$ for both local ULIRGs 
(e.g. Howell et al. 2010)
and z$>$2 submm-luminous galaxies.
On the other hand, it appears that the UV provides a better estimate 
(closer to the measured $L_{\rm IR}$) for the averaged BX$/$BMs.

Finally, it is instructive to look at variations of the obscuration of
these UV-selected galaxies. For this purpose we examine the bolometric
luminosity (defined as the sum of the IR and UV luminosities) as a
function of obscuration (approximated by the ratio of IR-to-UV
luminosity) for LBGs and BX$/$BMs. For comparison we include
submm-luminous and UGR-selected $z\sim$2 galaxies (from Reddy et
al. 2006 and references therein). The resulting plot is shown in
Fig. 5. The straight line indicates the correlation found by Reddy et
al. (2006, 2010) for $z\sim$2 UGR-selected galaxies. The averaged
$z\sim$3 LBG and the two individually detected ones appear to follow
the relation defined for the z$\sim$2 galaxies. In terms of
luminosities, both averaged LBGs and BX$/$BMs have similar $L_{UV}$
(few $\times$ 10$^{10}L_{\odot}$) but LBGs have higher $L_{\rm IR}$
and thus higher $L_{FIR}/L_{UV}$ ratio.  Since it is
well established that obscuration decreases with increasing redshift
(Reddy et al. 2006, 2010, Adelberger and Steidel 2000), the difference
in the $L_{FIR}/L_{UV}$ ratio must be attributed to different
causes. While selection effects are likely to play a role 
(see section 2.2) we argue that possible differences in morphologies, dust 
distribution and extent of star-forming regions are also likely to 
contribute. Morphological studies of UV-selected $z\sim$2 and z$\sim$3 
galaxies in the GOODS-N field find few differences between the two samples
(Law et al. 2007) although, dustier galaxies (as evidenced by E(B-V))
were found to show more nebulous UV morphologies. 
Finally, since MIPS-detected LBGs have ULIRG-like luminosities (section 3.1) 
it is possible that their UV and IR emission originates in different 
regions (as observed in local ULIRGs e.g. Wang et al 2004, see also Huang 
et al. 2007) and thus could account for the higher $L_{IR}/L_{UV}$ ratio 
observed.

\begin{figure}
\includegraphics[width=70mm, angle=-90]{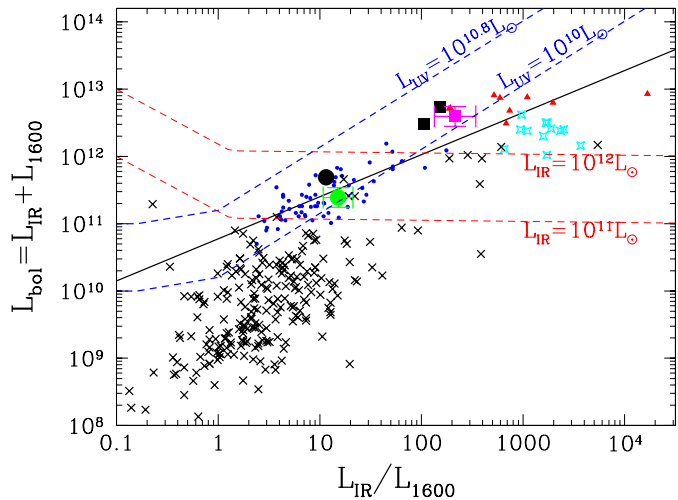} 
\caption{Bolometric luminosity, approximated as the sum of the IR and UV
luminosities, vs. IR-to-UV luminosity ratio (dust obscuration). 
Small blue circles are $z\sim$2
spectroscopically confirmed BX$/$BMs (from Reddy et al. 2006), red
triangles represent submm-luminous galaxies, black crosses are local
normal (Bell et al. 2003) and starbursts (Brandl et al. 2006) galaxies and
cyan stars are ULIRGs.
The large black circle corresponds to BM1326 while the
large green circle is the average BX$/$BM. Magenta and black squares
are the average LBGs and the two LBGs with mm detections.  The
solid line indicates the best-fit linear relation for
spectroscopically confirmed {\it UGR}-galaxies detected at 24 $\mu$m (from
Reddy et al. 2006). The red
and blue dashed lines are lines of constant $L_{\rm UV}$ and $L_{\rm IR}$
luminosity. The errorbars for the stacked LBG (magenta) and BX/BM
(green) values have been magnified for clarity.}
\end{figure}

\section*{Acknowledgments}

SPIRE has been developed by a consortium of institutes led by Cardiff
Univ. (UK) and including Univ. Lethbridge (Canada); NAOC (China); CEA,
LAM (France); IFSI, Univ. Padua (Italy); IAC (Spain); Stockholm
Observatory (Sweden); Imperial College London, RAL, UCL-MSSL, UKATC,
Univ. Sussex (UK); Caltech, JPL, NHSC, Univ. Colorado (USA). This
development has been supported by national funding agencies: CSA
(Canada); NAOC (China); CEA, CNES, CNRS (France); ASI (Italy); MCINN
(Spain); SNSB (Sweden); STFC (UK); and NASA (USA). The
data presented in this paper will be released through the Herschel
Database in Marseille HeDaM (hedam.oamp.fr$/$HerMES)

\end{document}